\documentclass[sn-mathphys,Numbered]{sn-jnl}

\usepackage{graphicx}%
\usepackage{multirow}%
\usepackage{amsmath,amssymb,amsfonts}%
\usepackage{amsthm}%
\usepackage{mathrsfs}%
\usepackage[title]{appendix}%
\usepackage{xcolor}%
\usepackage{textcomp}%
\usepackage{manyfoot}%
\usepackage{booktabs}%
\usepackage{algorithm}%
\usepackage{algorithmicx}%
\usepackage{algpseudocode}%
\usepackage{listings}%
\usepackage{adjustbox}
\usepackage{mathdots}
\usepackage{derivative}
\usepackage{array}

\theoremstyle{thmstyleone}%
\theoremstyle{thmstyletwo}%

\theoremstyle{thmstylethree}%

\begin{document}

\title[Article Title]{
  \begin{center}
    \textbf{Reaching the optical propagation limit\\ in temporal analog computing}
  \end{center}
}

\author*{\fnm{Zeki} \sur{Hayran}}\email{z.hayran@imperial.ac.uk}

\affil{The Blackett Laboratory, Department of Physics, Imperial College London, London SW7 2AZ, United Kingdom}

\abstract{
A central goal of optical computing is to perform calculations on the timescale of light propagation. Yet many analog photonic solvers require feedback, storage or field build-up before the answer becomes available, introducing additional latency that limits real-time operation. Here we introduce the concept of two-time modulation for temporal analog computing, in which the material response is independently modulated along two temporal directions, allowing the computational operator to be constructed continuously as the waveform propagates. The solution can therefore form during optical transit, without an additional solution-formation timescale. This form of temporal control allows identical copies of a waveform separated only in time to reach entirely different target outputs within a single spatial channel, and enables nonlocal integral-equation solving in a single passage. The resulting framework brings compact, programmable, real-time analog computation within reach for ultrafast optical information processing.}

\keywords{}

\maketitle
 
\section{Introduction}\label{sec1}

The growing energy and latency costs of moving and processing information have intensified the search for computing architectures that complement conventional digital hardware \cite{mcmahon2023physics,zangeneh2021analogue}. Optical computing offers one such route by mapping mathematical operations directly onto the propagation, interference and interaction of light \cite{solli2015analog,silva2014performing}. The large bandwidth and many simultaneously accessible degrees of freedom of optical fields have enabled processors for tasks ranging from linear transformations and signal processing to the solution of differential and integral equations \cite{silva2014performing,mohammadi2019inverse,cordaro2023solving,fu2024reconfigurable}, with recent nanophotonic platforms extending analog processing directly to ultrafast temporal and spatiotemporal signals \cite{esfahani2024tailoring,cotrufo2024temporal,huang2026experimental}. A particularly compelling prospect is to make the computation itself occur on the timescale of optical propagation \cite{mcmahon2023physics}.

\begin{figure}
 \centering
 \begin{adjustbox}{width=0.7\linewidth, center}
  \includegraphics[keepaspectratio]{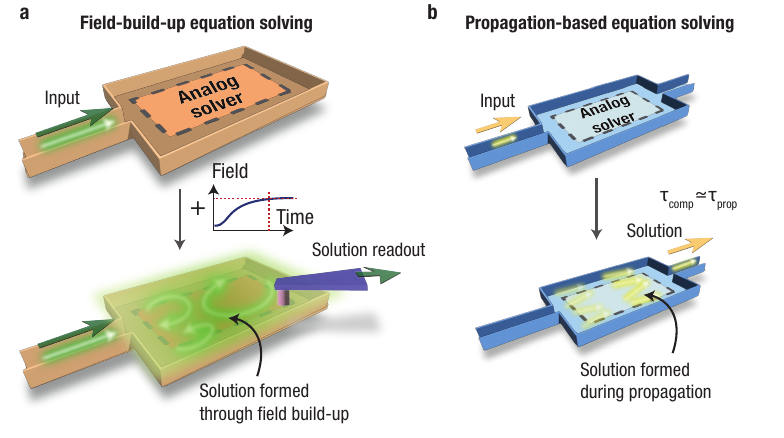}
 \end{adjustbox}
\caption{
\textbf{Field-build-up and propagation-based equation solving.}
\textbf{a,} In field-build-up approaches, the input excites a computing structure in which the solution is formed through development of a global field response before readout.
\textbf{b,} In propagation-based equation solving, the input is transformed continuously as it traverses the analog processor, so that the solution is formed during propagation and emerges directly at the output. With no separate solution-build-up stage, the intrinsic computation time approaches the optical transit time through the processor, \(\tau_{\mathrm{comp}}\simeq\tau_{\mathrm{prop}}\).
}
 \label{fig1}
\end{figure}

Yet the computational latency of an analog solver is set not by optical transit alone, but by when the solution becomes physically available. A broad class of photonic processors constructs the required response through resonant storage or feedback \cite{ferrera2010chip,hou2017optical,camacho2021single,cordaro2023solving,fu2024reconfigurable}; in closed-loop matrix inversion, for example, the solution is encoded in the steady-state response of the feedback network \cite{tzarouchis2025programmable}. The optical field may therefore traverse the computing structure while the computational state is still forming, introducing a solution-formation timescale beyond optical transit (Fig.~\ref{fig1}a). Reaching the optical propagation limit requires removing this additional timescale, so that the solution is formed during transit rather than after feedback, recurrence or field build-up.

One route to the optical propagation limit is to make propagation itself perform the computation, so that the output is formed as the field traverses the system. We refer to this as propagation-based optical computing (Fig.~\ref{fig1}b). Fourier optics provides a familiar example, where propagation through a lens system directly produces the Fourier-transformed field. More generally, propagation can implement complex optical transformations. Free-space diffractive processors and programmable multimode waveguides have used this principle for inference and general linear transformations \cite{lin2018alloptical,kulce2021alloptical,hu2024diffractive,onodera2026arbitrary}. Spatial analog processors have similarly embedded mathematical operations directly into scattering \cite{silva2014performing,koufidis2025chirality}, including equation-solving and nonlocal transformations in which each output depends on a range of input positions \cite{goh2022nonlocal,li2025spatial,kiani2026nonreciprocal}.

This raises a natural question of whether propagation-based computing can be carried out directly in waveform time rather than across spatial coordinates. Such a formulation would allow the computational function to remain encoded in a temporal waveform within a single spatial channel. Moving the function into time, however, changes the geometry of the computation. In spatial analog computing, the function is defined across coordinates transverse to propagation, leaving propagation as an independent direction along which the required couplings can be built. In temporal analog computing, the function is instead defined across waveform time. A general nonlocal temporal operator must therefore couple different waveform times while the transformation changes independently as propagation proceeds. A modulation specified only in waveform time cannot provide both roles. Propagation-based temporal equation solving therefore requires \textit{two temporal coordinates in the propagation description}, one to carry the function across the waveform and another to describe how the operation evolves as the waveform propagates.

Here, we introduce these two coordinates through a two-time representation of Maxwell propagation, which allows temporal input--output operators to be constructed continuously as light traverses the medium. We first use this approach to realize independently specified waveform transformations and then synthesize the nonlocal inverse operator of a Fredholm integral equation. Full-wave simulations verify the evolution from the encoded input to the computed solution, with the computation completed over the optical transit through the programmed medium.

\section{Two-time propagation for temporal operator synthesis}\label{sec2}

\begin{figure}
 \centering
 \begin{adjustbox}{width=1.0\linewidth, center}
  \includegraphics[keepaspectratio]{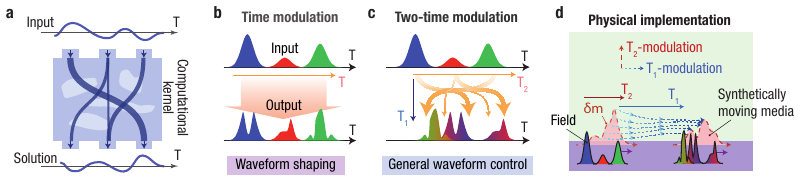}
 \end{adjustbox}
\caption{
\textbf{From temporal waveform shaping to temporal operator synthesis.}
\textbf{a,} A temporal waveform carries the input function while propagation transforms it into the solution of the target mathematical problem.
\textbf{b,} A modulation specified along a single waveform-time coordinate \(T\) can reshape the temporal profile but cannot independently vary the operation as propagation proceeds.
\textbf{c,} Two-time modulation separates waveform time \(T_2\), on which the computational function is defined, from evolution time \(T_1\), which orders the transformations applied during propagation. A material response structured over \(T_1\) and \(T_2\) can therefore synthesize a temporal input--output operator progressively through the medium.
\textbf{d,} Synthetic motion provides a physical route to two-time modulation. A material modulation travelling at the reference velocity follows the signal along \(T_2\), while independent variation of its profile along propagation supplies the \(T_1\) dependence.
}
 \label{fig2}
\end{figure}

A temporal waveform can carry the function on which a computation acts, while propagation transforms that function into the target solution (Fig.~\ref{fig2}a). Conventional temporal modulation can vary the material response across waveform time and thereby reshape the temporal profile, but it cannot independently specify how the operation changes as propagation proceeds (Fig.~\ref{fig2}b). We therefore separate these roles into waveform time \(T_2\), which carries the computational function, and evolution time \(T_1\), which orders its transformation during propagation (Fig.~\ref{fig2}c).

To formulate these two temporal roles, we express progression through the medium as elapsed time along a reference trajectory of velocity \(v_r\). The laboratory coordinates \(z\) and \(t\) are transformed according to
\begin{equation}
T_1=\beta_r(z-z_0),\qquad
T_2=t-\beta_r(z-z_0),\qquad
\beta_r=v_r^{-1},
\label{eq:two_time_coordinates}
\end{equation}
where \(z_0\) denotes the input plane. The inverse mapping is
\begin{equation}
z=z_0+v_rT_1,\qquad
t=T_1+T_2.
\label{eq:two_time_inverse}
\end{equation}
Thus \(T_1\) measures the elapsed propagation time through the medium, whereas \(T_2\) gives the time within the waveform relative to the same reference trajectory. These are simply new coordinates for the ordinary variables \(z\) and \(t\); no additional physical time dimension is introduced. If \(v_r\) is chosen to equal the pulse group velocity, the pulse centre remains fixed in \(T_2\), and propagation through the medium is described by increasing \(T_1\).

The derivative transformation,
\begin{equation}
\partial_z=\beta_r(\partial_{T_1}-\partial_{T_2}),
\qquad
\partial_t=\partial_{T_2},
\end{equation}
maps Maxwell equations exactly onto the \((T_1,T_2)\) plane before any envelope or forward-propagation approximation. For a stationary reference medium, a plane wave with propagation constant \(k(\omega)\) then acquires the two-time dispersion relation
\begin{equation}
\Omega_1(\omega)=\omega-v_r k(\omega),
\label{eq:two_time_dispersion}
\end{equation}
where \(\Omega_1\) is the frequency conjugate to \(T_1\).

We choose the reference velocity to match the group velocity at the carrier frequency, \(v_r=v_g(\omega_0)\). The total advance along the evolution coordinate through a medium of length \(L\) is then
\begin{equation}
\Delta T_1=\beta_r L=\frac{L}{v_g(\omega_0)}
\equiv\tau_{\mathrm{prop}},
\label{eq:T1_propagation_time}
\end{equation}
so the span of the processor along \(T_1\) is its optical propagation time. Selecting the forward-propagating branch and expanding Eq.~\ref{eq:two_time_dispersion} about \(\omega_0\) give, in the narrowband quadratic-dispersion limit,
\begin{equation}
i\partial_{T_1}A(T_1,T_2)
=
-\frac{B_0}{2}\partial_{T_2}^{2}A(T_1,T_2)
+
V(T_1,T_2)A(T_1,T_2),
\label{eq:two_time_envelope}
\end{equation}
where \(A\) is the complex field envelope, \(B_0=-v_r\beta_2(\omega_0)\), and \(V(T_1,T_2)\) is the effective local material modulation. Equation~\ref{eq:two_time_envelope} provides the scalar description used below to establish general waveform control. The Maxwell transformation, causal dispersive constitutive response \cite{solis2021functional,hayran2022hbar,koutserimpas2024time}, and forward-wave reduction leading to this limit are developed in Supplementary Sections~S1--S3. The Fredholm implementation below retains the time-dependent dispersive Drude response, as described in Methods.

Equation~\ref{eq:two_time_envelope} also shows how a temporal operator can be assembled during propagation. At each \(T_1\), the waveform evolves under the generator
\begin{equation}
\hat{H}(T_1)
=
-\frac{B_0}{2}\partial_{T_2}^{2}
+
V(T_1,T_2),
\label{eq:two_time_hamiltonian}
\end{equation}
which specifies the temporal transformation applied at that stage of propagation. If \(V\) is independent of \(T_1\), the same generator acts throughout the medium. Two-time modulation instead allows the generator to change as the waveform propagates, so that different temporal operations are applied successively. These operations generally do not commute, \([\hat{H}(T_1),\hat{H}(T_1')]\neq0\), and their order therefore contributes to the final transformation. The complete input--output operator is
\begin{equation}
\hat{\mathcal{U}}(T_{1,f},T_{1,i})
=
\mathcal{T}_{1}
\exp\left[
-i\int_{T_{1,i}}^{T_{1,f}}
\hat{H}(T_1)\,dT_1
\right],
\label{eq:ordered_propagator}
\end{equation}
where \(\mathcal{T}_{1}\) orders the generators along \(T_1\). The corresponding Magnus expansion is developed in Supplementary Section~S4.

The resulting propagation operator can be written as a temporal transfer kernel,
\begin{equation}
S(T_2,T_2')
=
\left\langle T_2
\middle|
\hat{\mathcal{U}}(T_{1,f},T_{1,i})
\middle|
T_2'
\right\rangle,
\qquad
A_{\mathrm{out}}(T_2)
=
\int_{\mathcal{T}}
S(T_2,T_2')A_{\mathrm{in}}(T_2')\,dT_2'.
\label{eq:temporal_transfer_kernel}
\end{equation}
The kernel \(S(T_2,T_2')\) gives the contribution of the input field at waveform time \(T_2'\) to the output field at waveform time \(T_2\). Dispersion couples different waveform times, while the two-time modulation changes these couplings as propagation proceeds. The output at a given \(T_2\) can therefore depend on the input over a range of \(T_2'\), producing a nonlocal temporal operator. This is the \textit{temporal analogue of a spatially nonlocal optical kernel}, with coupling between different positions replaced by coupling between different times within a single travelling waveform \cite{goh2022nonlocal,li2025spatial}. The independent evolution coordinate \(T_1\) allows this kernel to be assembled progressively during propagation. In the Fredholm calculation below, inverse design is used to make the physical propagation operator approximate the inverse operator required by the target equation.

To realize two-time modulation physically, the material response must follow the travelling waveform while its temporal profile changes along propagation. Synthetic motion provides a route to this behaviour, with spatiotemporal shaping producing a material modulation that travels at a specified apparent velocity \cite{huidobro2019fresnel,galiffi2022photonics,harwood2025space,hayran2026space,harwood2026programmable}. Matching this velocity to the reference velocity \(v_r\) makes the modulation co-moving with the signal along \(T_2\), while variation of its profile along propagation supplies the independent \(T_1\) dependence. The corresponding laboratory-frame response is
\begin{equation}
m(z,t)
=
m_{2t}\!\left[
\beta_r(z-z_0),
t-\beta_r(z-z_0)
\right],
\label{eq:synthetic_motion_mapping}
\end{equation}
where \(m\) denotes the physical material parameter being modulated. The second argument follows the waveform coordinate \(T_2\), while the first argument, \(\beta_r(z-z_0)=T_1\), describes how the co-moving temporal profile changes during propagation. If the response is independent of the first argument, Eq.~\ref{eq:synthetic_motion_mapping} reduces to the synthetically moving form \(m(z,t)=m_0[t-\beta_r(z-z_0)]\), or equivalently \(m_{2t}(T_1,T_2)=m_0(T_2)\). Two-time modulation extends this response to the general form \(m_{2t}(T_1,T_2)\), in which the co-moving temporal profile itself evolves along \(T_1\), as illustrated in Fig.~\ref{fig2}d. In the scalar description of Eq.~\ref{eq:two_time_envelope}, the material response is represented by \(V(T_1,T_2)\); in the Maxwell--Drude Fredholm implementation below, the programmed parameter is the plasma frequency, \(m=\omega_p\). Velocity mismatch and finite temporal acceptance are treated in Supplementary Section~S5.

\section{Propagation-based temporal analog computing}\label{sec3}

\begin{figure} [H]
 \centering
 \begin{adjustbox}{width=0.9\linewidth, center}
  \includegraphics[keepaspectratio]{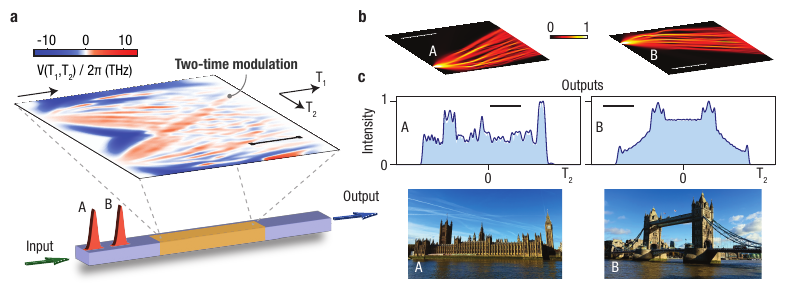}
 \end{adjustbox}
\caption{
\textbf{Programmable waveform transformations by two-time modulation.}
\textbf{a,} Two identical input pulses, \(A\) and \(B\), with the same temporal profile and spectral intensity but different launch times, propagate through the same inverse-designed modulation \(V(T_1,T_2)\).
\textbf{b,} Evolution of the normalized field magnitude, \(|A(T_1,T_2)|\), for the two time-shifted inputs.
\textbf{c,} Normalized output intensities, \(|A(T_2)|^2\), and the corresponding target profiles associated with the Palace of Westminster and Tower Bridge. The intensity-profile fidelities are \(0.982\) and \(0.980\), respectively, where unity corresponds to exact agreement with the target intensity profile (see Methods section). Scale bars in \textbf{a--c}, 500 fs.
}
 \label{fig3}
\end{figure}

Before using two-time modulation for equation solving, we first demonstrate general temporal control within a single spatial channel. Figure~\ref{fig3}a shows two pulses, \(A\) and \(B\), with identical temporal profiles and spectral intensities but different launch times, propagating through the same inverse-designed modulation \(V(T_1,T_2)\). A temporal translation changes only the spectral phase of the input, but shifts the waveform along \(T_2\) within the two-time modulation. Because \(\hat{H}(T_1)\) is structured across \(T_2\) and evolves along \(T_1\), the same ordered sequence of propagation generators acts differently on the two shifted states, producing the distinct field evolutions shown in Fig.~\ref{fig3}b. At the output, the two pulses reproduce independently specified intensity profiles associated with the Palace of Westminster and Tower Bridge, with intensity-profile fidelities of \(0.982\) and \(0.980\), respectively, where unity corresponds to exact agreement with the target intensity profile (see Methods section; Fig.~\ref{fig3}c). Two time-shifted copies of the same waveform can thus undergo different target transformations in the same programmed medium and within a single spatial channel. Details of the inverse design and fidelity measure are given in Methods and Supplementary Section~S6.

We next use this control to implement an equation-solving operator. We consider a Fredholm equation of the second kind, a class of integral equations previously used as a benchmark for wave-based analog equation solving \cite{mohammadi2019inverse,cordaro2023solving},
\begin{equation}
u(x)
=
f(x)
+
\int_{-7.2}^{7.2}
\kappa(x,y)u(y)\,dy,
\label{eq:fredholm}
\end{equation}
where \(x=T_2/T_s\), \(y=T_2^{\prime}/T_s\) and \(T_s=125~\mathrm{fs}\). The complex kernel \(\kappa(x,y)\) couples different parts of the temporal function, so that the solution at one waveform time depends on the function over a range of other waveform times. Since \(u\) also appears inside the integral, solving the equation requires the inverse operator
\begin{equation}
\hat{S}
=
\left(
\hat{I}-\hat{K}
\right)^{-1},
\label{eq:fredholm_inverse}
\end{equation}
such that \(u=\hat{S}f\). The optical task is therefore to implement this complete inverse transformation on the input waveform. We represent the equation on five orthonormal localized temporal modes, with the continuous kernel defining the corresponding \(5\times5\) operator as described in Methods and Supplementary Section~S7. The modal amplitudes are carried together by a single temporal waveform, while the off-diagonal elements describe coupling between different temporal modes. The input function \(f(x)\) is encoded in the complex field envelope at the entrance of the computing region, \(A(0,T_2)\), with \(x=T_2/T_s\), while the output envelope \(A(\Delta T_1,T_2)\) represents the solution \(u(x)\).

\begin{figure}
 \centering
 \begin{adjustbox}{width=0.8\linewidth, center}
  \includegraphics[keepaspectratio]{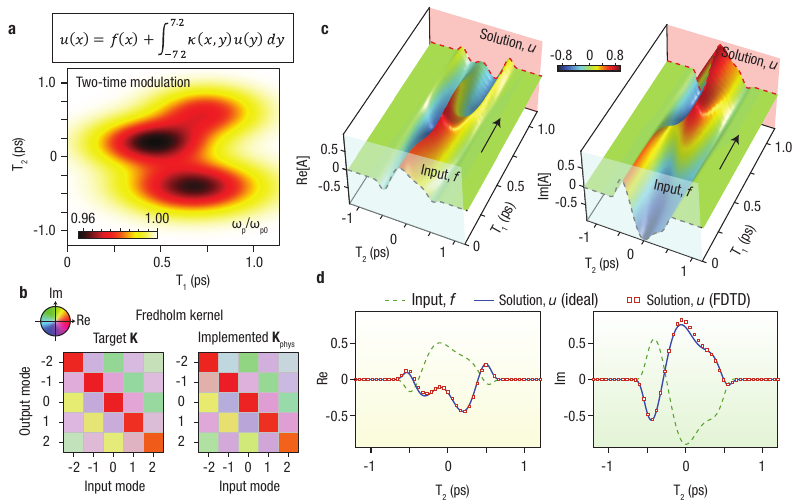}
 \end{adjustbox}
\caption{
\textbf{Propagation-based solution of a nonlocal Fredholm equation.}
\textbf{a,} Fredholm equation considered and inverse-designed normalized plasma-frequency modulation, \(\omega_p(T_1,T_2)/\omega_{p0}\), with a maximum fractional modulation of \(4.49\%\).
\textbf{b,} Complex modal representations of the target Fredholm kernel \(\mathbf{K}\) and the implemented kernel \(\mathbf{K}_{\mathrm{phys}}\) reconstructed from the complete five-mode propagation operator. Their normalized complex overlap is \(\mathcal{O}_K=0.9985\) (see Methods).
\textbf{c,} Real and imaginary parts of the complex field envelope \(A(T_1,T_2)\) obtained from full-wave Maxwell--Drude FDTD and projected onto the five-mode computational space. The input envelope encodes \(f\), while the output envelope represents the solution \(u\).
\textbf{d,} Input waveform encoding \(f\), ideal Fredholm solution \(u\), and corresponding projected full-wave FDTD solution for the real and imaginary field quadratures.
}
 \label{fig4}
\end{figure}

To realize this transformation, we program the plasma frequency of a dispersive Drude medium over \(T_1\) and \(T_2\) and inverse-design its propagation operator towards \(\hat{S}\). The calculation includes the change in local material dispersion produced by the plasma-frequency modulation, together with the accessible modulation range. The optimized profile \(\omega_p(T_1,T_2)/\omega_{p0}\) is shown in Fig.~\ref{fig4}a and reaches a maximum fractional modulation of \(4.49\%\). At each position along \(T_1\), the variation of \(\omega_p\) across \(T_2\) defines a temporally structured dispersive response. This response changes continuously along propagation, so that the successive transformations experienced by the waveform combine to produce the target inverse operator.

To characterize the implemented Fredholm operator, we reconstruct the complete five-mode propagation matrix. Each basis mode is propagated through the optimized medium and the corresponding output is projected onto the same temporal basis. This gives the physical propagation matrix \(\mathbf{S}_{\mathrm{phys}}\), designed to satisfy
\begin{equation}
\mathbf{S}_{\mathrm{phys}}
\simeq
\left(
\mathbf{I}-\mathbf{K}
\right)^{-1},
\qquad
\mathbf{K}_{\mathrm{phys}}
=
\mathbf{I}
-
\mathbf{S}_{\mathrm{phys}}^{-1}.
\label{eq:implemented_fredholm}
\end{equation}
Figure~\ref{fig4}b compares the target Fredholm kernel \(\mathbf{K}\) with the kernel \(\mathbf{K}_{\mathrm{phys}}\) obtained from this propagation operator. Both the diagonal response and the couplings between different temporal modes are reproduced across the complete \(5\times5\) operator, with a normalized complex overlap of \(\mathcal{O}_K=0.9985\) between the target and implemented kernels (see Methods). The material model, finite-mode projection and inverse-design procedure are described in Methods and Supplementary Sections~S7 and S8.

We finally test the optimized medium using full-wave Maxwell--Drude simulations for an input waveform composed of a nontrivial superposition of the computational modes. Figure~\ref{fig4}c shows the real and imaginary parts of the complex field envelope, projected onto the five-mode space, throughout the computing region. Both quadratures evolve continuously as the input waveform propagates through the programmed medium and approach the computed solution at the output. The output waveform is compared directly with the ideal Fredholm solution in Fig.~\ref{fig4}d. Using the modal coefficients measured at the FDTD entrance to calculate the corresponding ideal solution, the projected full-wave output reaches a normalized field overlap of \(0.9974\) with the ideal solution, where unity corresponds to identical normalized complex field structure up to an overall complex factor (Methods).

For the physical scaling considered here, the computing region is \(100~\mu\mathrm{m}\) long and has an optical transit time of approximately \(1.14~\mathrm{ps}\). The evolution shown in Fig.~\ref{fig4}c therefore takes place over the optical transit itself, with the Fredholm solution formed as the waveform reaches the output. The computation is completed within this propagation interval, giving \(\tau_{\mathrm{comp}}\simeq\tau_{\mathrm{prop}}\). Details of the full-wave validation are given in Methods and Supplementary Section~S9.

\section{Discussion}\label{sec4}

Two-time modulation allows the inverse operator required for temporal equation solving to be constructed during propagation, while the mathematical function remains encoded across the travelling waveform. The intrinsic solution-formation time is therefore set by the optical transit through the computing medium, according to the latency definition in Methods. This raises a further question: how short can that transit time itself be? The required temporal transformation must still be accumulated during propagation, so its implementation requires a minimum propagation distance and hence a \textit{minimum transit time}. This is analogous to spatial optical systems, where the nonlocality required by a target operation can impose a \textit{minimum device thickness} \cite{miller2023why}. Establishing the corresponding lower bound in time for temporal analog computing, including its dependence on temporal bandwidth, material response and operator complexity, will be the subject of future work.

A related question is how large a temporal computation can be carried by a given system. The Fredholm example uses five temporal modes, but the same formulation can operate on a larger temporal space. The number of accessible degrees of freedom is set by the available temporal time-bandwidth product and by how much structure can be programmed independently along propagation. Increasing these resources allows correspondingly larger operators and more complex integral kernels. The same temporal degrees of freedom could also form the nodes of an optical neural network, with individual temporal bins acting as network nodes and two-time modulation providing programmable mixing between them during propagation. This would provide a temporal counterpart to spatial optical networks in which diffraction or multimode propagation connects spatial degrees of freedom \cite{lin2018alloptical,onodera2026arbitrary}. Practical scaling will ultimately be constrained by material bandwidth, loss and modulation complexity, while the conditioning of the mathematical operator determines sensitivity to physical errors. In particular, inverses close to singularity require large singular-value amplification and will generally demand regularization or preconditioning. More general nonunitary transformations can also be accommodated through material loss or coupling between the computational space and auxiliary optical channels, as described in Supplementary Section~S4.

How these limits are reached in practice will depend on the physical platform. The essential ingredients are temporal mixing through dispersion, which plays the role of diffraction in the spatial analogue, and a material response that can be structured across the travelling waveform while varying independently as propagation proceeds. Space-time modulation and synthetic motion provide natural routes to such responses \cite{galiffi2022photonics,ciabattoni2025observation,harwood2025space,harwood2026programmable}. Recent experiments have demonstrated large and rapidly varying optical responses capable of producing synthetic motion and programmable space-time transformations \cite{harwood2025space,harwood2026programmable}. Extending these approaches to a distributed interaction would allow the two-time operator to be programmed directly in physical space-time. The achievable transformation would then be set by the available dispersion, modulation timescale and strength.

Beyond analog computing, two-time modulation introduces a broader class of temporal photonic systems in which the material response can be structured over an effective \(T_1,T_2\) plane. This opens the possibility of constructing temporal analogues of concepts that ordinarily require two spatial coordinates, including lattices, interfaces and defects, with their structure evolving during propagation. Periodicity along both temporal coordinates could further generalize photonic time crystals \cite{asgari2024theory} from modulation along a single temporal direction to band structures defined across a two-coordinate temporal geometry. More broadly, the two-time plane opens a route to temporal analogues of genuinely two-dimensional wave phenomena, extending temporal photonics beyond the physics accessible along a single time coordinate.

\section{Methods}\label{sec:methods}

\subsection{Temporal waveform inverse design}

The waveform transformations in Fig.~\ref{fig3} are calculated using Eq.~\ref{eq:two_time_envelope} with \(B_0=-10~\mathrm{fs}\) and \(|V|/(2\pi)\leq15~\mathrm{THz}\). The two incident fields are identical normalized Gaussian pulses with an intensity RMS width of \(8~\mathrm{fs}\), centred at \(T_2=-0.9~\mathrm{ps}\) and \(T_2=+0.9~\mathrm{ps}\). Their temporal displacement changes only the spectral phase, so the two inputs have identical temporal profiles and spectral intensities.

The real potential \(V(T_1,T_2)\) is inverse-designed simultaneously for the two target output intensities. If \(p_j(T_2)\) denotes a normalized target intensity and \(I_j(T_2)\) the calculated output intensity, the fidelity is
\begin{equation}
S_j=
\left[
\sum_n\sqrt{p_{j,n}I_{j,n}}
\right]^2.
\label{eq:methods_fidelity}
\end{equation}
Here \(S_j=1\) corresponds to exact agreement between the normalized calculated and target intensity profiles. The optimization balances the fidelities of both transformations, with gradients calculated from the adjoint of the discrete propagation operator and the parameters updated using Adam \cite{kingma2015adam}. The parameterization, numerical discretization, regularization and target preparation are described in Supplementary Section~S6.

\subsection{Fredholm operator and physics-constrained inverse design}

For the Fredholm calculation, \(x=T_2/T_s\), \(y=T_2'/T_s\), \(s=(x+y)/2\), \(d=x-y\) and \(T_s=125~\mathrm{fs}\). The analytical kernel is
\begin{equation}
\kappa(x,y)
=
C_0e^{-p_0d^2+i\beta_0d}
+
e^{-pd^2+i(\beta d+qs)}
\left[
C_1+C_2\cos(as+b)
\right],
\label{eq:methods_fredholm_kernel}
\end{equation}
with
\begin{equation}
C_0=-0.2-2.2i,\quad
C_1=3.1+4.0i,\quad
C_2=0.1-0.7i,
\end{equation}
and
\begin{equation}
p_0=1.3,\quad
\beta_0=0.5,\quad
p=4.6,\quad
\beta=0.7,\quad
q=0,\quad
a=1.3,\quad
b=-2.3.
\end{equation}
The equation is defined over \(-7.2\leq x,y\leq7.2\). Its projection onto five localized temporal modes with nominal frequency offsets \(\{-1.6,-0.8,0,0.8,1.6\}~\mathrm{THz}\), orthonormalized using a symmetric L\"owdin transformation \cite{lowdin1950nonorthogonality}, defines the matrix \(\mathbf K\) and target solution operator
\begin{equation}
\mathbf S_{\mathrm{target}}
=
\left(
\mathbf I-\mathbf K
\right)^{-1}.
\label{eq:methods_target_operator}
\end{equation}

The physical operator is implemented in a \(100~\mu\mathrm{m}\) dispersive Drude medium with \(\lambda_0=7.75~\mu\mathrm{m}\), \(\epsilon_\infty=3.8\), \(\omega_{p0}/\omega_0=1.6\) and \(\gamma/(2\pi)=60~\mathrm{GHz}\). The plasma frequency is programmed over the two-time plane with \(-0.05\leq\Delta\omega_p/\omega_{p0}\leq0\). Each candidate material profile is therefore evaluated through the corresponding dispersive propagation dynamics, and the projected five-mode propagation matrix is optimized towards \(\mathbf S_{\mathrm{target}}\). The normalized complex overlap between the target and implemented kernels is defined as
\begin{equation}
\mathcal{O}_K
=
\frac{
\left|
\mathrm{Tr}
\left(
\mathbf{K}^{\dagger}\mathbf{K}_{\mathrm{phys}}
\right)
\right|
}{
\|\mathbf{K}\|_F
\|\mathbf{K}_{\mathrm{phys}}\|_F
}.
\label{eq:methods_kernel_overlap}
\end{equation}
Here \(\mathcal{O}_K=1\) corresponds to identical normalized complex matrix structure up to an overall complex factor. The corresponding relative Frobenius errors of the kernel and solution operator are reported in Supplementary Section~S8. The modal projection, dispersive propagation model and optimization procedure are described in Supplementary Sections~S7 and S8.

\subsection{Full-wave validation and computational latency}

The inverse-designed plasma-frequency profile is validated independently using one-dimensional Maxwell--Drude finite-difference time-domain simulations with an auxiliary-current formulation. The incident waveform is synthesized from the same five temporal modes used to define the Fredholm problem. At each position, the optical field is converted to a complex envelope in the co-moving coordinate \(T_2\) and projected onto the five-mode computational space. The ideal Fredholm solution used for comparison is evaluated from the modal coefficients of the field measured at the FDTD entrance. The relative field error and normalized overlap are defined as
\begin{equation}
\epsilon(a,b)
=
\frac{\|a-b\|_2}{\|b\|_2},
\qquad
\mathcal{O}(a,b)
=
\frac{|a^\dagger b|}{\|a\|_2\|b\|_2}.
\label{eq:methods_fdtd_metrics}
\end{equation}
Here \(\epsilon=0\) corresponds to exact complex-field agreement, while \(\mathcal{O}=1\) corresponds to identical normalized complex field structure up to an overall complex factor. The FDTD discretization and envelope extraction are described in Supplementary Section~S9.

For the group-velocity-matched coordinates, the optical propagation time through the computing region is
\begin{equation}
\tau_{\mathrm{prop}}
=
\frac{L}{v_g(\omega_0)}.
\end{equation}
We define the intrinsic excess solution-formation time as
\begin{equation}
\tau_{\mathrm{excess}}
=
\tau_{\mathrm{comp}}
-
\tau_{\mathrm{prop}}.
\end{equation}
Because the target computational operator is implemented by forward propagation itself, no subsequent recurrence, settling or feedback process is required after the waveform reaches the output plane. The ideal intrinsic latency therefore satisfies \(\tau_{\mathrm{excess}}=0\) and \(\tau_{\mathrm{comp}}=\tau_{\mathrm{prop}}\). If the interval is instead measured from arrival of the first input sample to departure of the last output sample for a waveform of duration \(T_{\mathrm{sig}}\), the corresponding wall-clock interval is \(\tau_{\mathrm{prop}}+T_{\mathrm{sig}}\).

\backmatter

\bmhead{Supplementary information}
Supplementary Sections S1--S9 provide the exact two-time Maxwell formulation, dispersive constitutive response, forward-wave and envelope limits, ordered temporal operator synthesis, synthetic-motion realization, waveform inverse design, Fredholm modal projection, physics-constrained Maxwell--Drude inverse design and full-wave validation.


\section*{Declarations}

\bmhead{Funding}
The author acknowledges support from the Engineering and Physical Sciences Research Council (EPSRC) under grant EP/Y015673/1.

\bmhead{Conflict of interest/Competing interests}
The author declares no competing interests.

\bmhead{Availability of data and materials}
The author confirms that all relevant data are included in the paper and/or its Supplementary Information files.

\bmhead{Code availability}
The code used to generate the results is available from the corresponding author upon reasonable request.

\bibliography{bibliography}

@article{mcmahon2023physics,
  title={The physics of optical computing},
  author={McMahon, Peter L.},
  journal={Nature Reviews Physics},
  volume={5},
  pages={717--734},
  year={2023},
  doi={10.1038/s42254-023-00645-5}
}

@article{miller2023why,
  title={Why optics needs thickness},
  author={Miller, David A. B.},
  journal={Science},
  volume={379},
  number={6627},
  pages={41--45},
  year={2023},
  doi={10.1126/science.ade3395}
}

@article{esfahani2024tailoring,
  title={Tailoring space-time nonlocality for event-based image processing metasurfaces},
  author={Esfahani, Sedigheh and Cotrufo, Michele and Al{\`u}, Andrea},
  journal={Physical Review Letters},
  volume={133},
  number={6},
  pages={063801},
  year={2024},
  doi={10.1103/PhysRevLett.133.063801}
}

@article{kiani2026nonreciprocal,
  title={Nonreciprocal nonlocal metasurface for multifunctional image processor},
  author={Kiani, Mehdi and Goh, Heedong and Al{\`u}, Andrea},
  journal={npj Metamaterials},
  volume={2},
  pages={7},
  year={2026},
  doi={10.1038/s44455-026-00018-9}
}

@article{zangeneh2021analogue,
  title={Analogue computing with metamaterials},
  author={Zangeneh-Nejad, Farzad and Sounas, Dimitrios L. and Al{\`u}, Andrea and Fleury, Romain},
  journal={Nature Reviews Materials},
  volume={6},
  pages={207--225},
  year={2021},
  doi={10.1038/s41578-020-00243-2}
}

@article{solli2015analog,
  title={Analog optical computing},
  author={Solli, Daniel R. and Jalali, Bahram},
  journal={Nature Photonics},
  volume={9},
  pages={704--706},
  year={2015},
  doi={10.1038/nphoton.2015.208}
}

@article{silva2014performing,
  title={Performing mathematical operations with metamaterials},
  author={Silva, Alexandre and Monticone, Francesco and Castaldi, Giuseppe and Galdi, Vincenzo and Al{\`u}, Andrea and Engheta, Nader},
  journal={Science},
  volume={343},
  number={6167},
  pages={160--163},
  year={2014},
  doi={10.1126/science.1242818}
}

@article{cotrufo2024temporal,
  title={Temporal signal processing with nonlocal optical metasurfaces},
  author={Cotrufo, Michele and Esfahani, Sedigheh and Korobkin, Dmitriy and Al{\`u}, Andrea},
  journal={npj Nanophotonics},
  volume={1},
  pages={39},
  year={2024},
  publisher={Nature Publishing Group},
  doi={10.1038/s44310-024-00039-0}
}

@article{huang2026experimental,
  title={Experimental demonstration of spatiotemporal analog computation in ultrafast optics},
  author={Huang, Junyi and Zhao, Dong and Shi, Jixuan and Zhang, Hongliang and Wang, Hengyi and Sun, Fang-Wen and Zhan, Qiwen and Zhu, Shiyao and Huang, Kun and Ruan, Zhichao},
  journal={Light: Science \& Applications},
  volume={15},
  pages={77},
  year={2026},
  publisher={Nature Publishing Group},
  doi={10.1038/s41377-025-02109-0}
}

@article{mohammadi2019inverse,
  title={Inverse-designed metastructures that solve equations},
  author={Mohammadi Estakhri, Nasim and Edwards, Brian and Engheta, Nader},
  journal={Science},
  volume={363},
  number={6433},
  pages={1333--1338},
  year={2019},
  doi={10.1126/science.aaw2498}
}

@article{ferrera2010chip,
  title={On-chip {CMOS}-compatible all-optical integrator},
  author={Ferrera, Marcello and Park, Youngmin and Razzari, Luca and Little, Brent E. and Chu, Sai T. and Morandotti, Roberto and Moss, David J. and Aza{\~n}a, Jos{\'e}},
  journal={Nature Communications},
  volume={1},
  pages={29},
  year={2010},
  doi={10.1038/ncomms1028}
}

@article{hou2017optical,
  title={Optical solver for a system of ordinary differential equations based on an external feedback assisted microring resonator},
  author={Hou, Jie and Dong, Jianji and Zhang, Xinliang},
  journal={Optics Letters},
  volume={42},
  number={12},
  pages={2310--2313},
  year={2017},
  doi={10.1364/OL.42.002310}
}

@article{camacho2021single,
  title={A single inverse-designed photonic structure that performs parallel computing},
  author={Camacho, Miguel and Edwards, Brian and Engheta, Nader},
  journal={Nature Communications},
  volume={12},
  pages={1466},
  year={2021},
  doi={10.1038/s41467-021-21664-9}
}

@article{cordaro2023solving,
  title={Solving integral equations in free space with inverse-designed ultrathin optical metagratings},
  author={Cordaro, Andrea and Edwards, Brian and Nikkhah, Vahid and Al{\`u}, Andrea and Engheta, Nader and Polman, Albert},
  journal={Nature Nanotechnology},
  volume={18},
  pages={365--372},
  year={2023},
  doi={10.1038/s41565-022-01297-9}
}

@article{fu2024reconfigurable,
  title={Reconfigurable metamaterial processing units that solve arbitrary linear calculus equations},
  author={Fu, Pengyu and Xu, Zimeng and Zhou, Tiankuang and Li, Hao and Wu, Jiamin and Dai, Qionghai and Li, Yue},
  journal={Nature Communications},
  volume={15},
  pages={6258},
  year={2024},
  doi={10.1038/s41467-024-50483-x}
}

@article{tzarouchis2025programmable,
  title={Programmable wave-based analog computing machine: a metastructure that designs metastructures},
  author={Tzarouchis, Dimitrios C. and Edwards, Brian and Engheta, Nader},
  journal={Nature Communications},
  volume={16},
  pages={908},
  year={2025},
  doi={10.1038/s41467-025-56019-1}
}

@article{goh2022nonlocal,
  title={Nonlocal scatterer for compact wave-based analog computing},
  author={Goh, Heedong and Al{\`u}, Andrea},
  journal={Physical Review Letters},
  volume={128},
  number={7},
  pages={073201},
  year={2022},
  doi={10.1103/PhysRevLett.128.073201}
}

@article{ashby2020temporal,
  title={Temporal mode transformations by sequential time and frequency phase modulation for applications in quantum information science},
  author={Ashby, James and Thiel, Val{\'e}rian and Allgaier, Markus and D'Ornellas, Peru and Davis, Alex O. C. and Smith, Brian J.},
  journal={Optics Express},
  volume={28},
  number={25},
  pages={38376--38389},
  year={2020},
  doi={10.1364/OE.410371}
}

@article{plansinis2016temporal,
  title={Temporal waveguides for optical pulses},
  author={Plansinis, Brent W. and Donaldson, William R. and Agrawal, Govind P.},
  journal={Journal of the Optical Society of America B},
  volume={33},
  number={6},
  pages={1112--1119},
  year={2016},
  doi={10.1364/JOSAB.33.001112}
}

@article{dong2023spatiotemporal,
  title={Spatiotemporal coupled-mode equations for arbitrary pulse transformation},
  author={Dong, Zhaohui and Chen, Xianfeng and Yuan, Luqi},
  journal={Physical Review Research},
  volume={5},
  number={4},
  pages={043150},
  year={2023},
  doi={10.1103/PhysRevResearch.5.043150}
}

@article{solis2021functional,
  title={Functional analysis of the polarization response in linear time-varying media: A generalization of the Kramers-Kronig relations},
  author={Sol{\'i}s, Diego M. and Engheta, Nader},
  journal={Physical Review B},
  volume={103},
  number={14},
  pages={144303},
  year={2021},
  publisher={American Physical Society},
  doi={10.1103/PhysRevB.103.144303}
}

@article{koutserimpas2024time,
  title={Time-varying media, dispersion, and the principle of causality},
  author={Koutserimpas, Theodoros T. and Monticone, Francesco},
  journal={Optical Materials Express},
  volume={14},
  number={5},
  pages={1222--1236},
  year={2024},
  publisher={Optica Publishing Group},
  doi={10.1364/OME.515957}
}

@article{huidobro2019fresnel,
  title={Fresnel drag in space-time-modulated metamaterials},
  author={Huidobro, Paloma A. and Galiffi, Emanuele and Guenneau, S{\'e}bastien and Craster, Richard V. and Pendry, J. B.},
  journal={Proceedings of the National Academy of Sciences},
  volume={116},
  number={50},
  pages={24943--24948},
  year={2019},
  publisher={National Academy of Sciences},
  doi={10.1073/pnas.1915027116}
}

@article{galiffi2022photonics,
  title={Photonics of time-varying media},
  author={Galiffi, Emanuele and Tirole, Romain and Yin, Shixiong and Li, Huanan and Vezzoli, Stefano and Huidobro, Paloma A. and Silveirinha, M{\'a}rio G. and Sapienza, Riccardo and Al{\`u}, Andrea and Pendry, J. B.},
  journal={Advanced Photonics},
  volume={4},
  number={1},
  pages={014002},
  year={2022},
  publisher={SPIE},
  doi={10.1117/1.AP.4.1.014002}
}

@article{harwood2025space,
  title={Space-time optical diffraction from synthetic motion},
  author={Harwood, A. C. and Vezzoli, S. and Raziman, T. V. and Hooper, C. and Tirole, R. and Wu, F. and Maier, S. A. and Pendry, J. B. and Horsley, S. A. R. and Sapienza, R.},
  journal={Nature Communications},
  volume={16},
  number={1},
  pages={5147},
  year={2025},
  publisher={Nature Publishing Group},
  doi={10.1038/s41467-025-60159-9}
}

@article{hu2024diffractive,
  title={Diffractive optical computing in free space},
  author={Hu, Jingtian and Mengu, Deniz and Tzarouchis, Dimitrios C. and Edwards, Brian and Engheta, Nader and Ozcan, Aydogan},
  journal={Nature Communications},
  volume={15},
  number={1},
  pages={1525},
  year={2024},
  publisher={Nature Publishing Group},
  doi={10.1038/s41467-024-45982-w}
}

@article{lin2018alloptical,
  title={All-optical machine learning using diffractive deep neural networks},
  author={Lin, Xing and Rivenson, Yair and Yardimci, Nezih T. and Veli, Muhammed and Luo, Yi and Jarrahi, Mona and Ozcan, Aydogan},
  journal={Science},
  volume={361},
  number={6406},
  pages={1004--1008},
  year={2018},
  publisher={American Association for the Advancement of Science},
  doi={10.1126/science.aat8084}
}

@article{li2025spatial,
  title={The spatial complexity of optical computing: toward space-efficient design},
  author={Li, Yandong and Monticone, Francesco},
  journal={Nature Communications},
  volume={16},
  pages={8588},
  year={2025},
  publisher={Nature Publishing Group},
  doi={10.1038/s41467-025-63453-8}
}

@article{onodera2026arbitrary,
  title={Arbitrary control over multimode wave propagation for machine learning},
  author={Onodera, Tatsuhiro and Stein, Martin M. and Ash, Benjamin A. and Sohoni, Mandar M. and Bosch, Melissa and Yanagimoto, Ryotatsu and Jankowski, Marc and McKenna, Timothy P. and Wang, Tianyu and Shvets, Gennady and Shcherbakov, Maxim R. and Wright, Logan G. and McMahon, Peter L.},
  journal={Nature Physics},
  volume={22},
  pages={164--171},
  year={2026},
  publisher={Nature Publishing Group},
  doi={10.1038/s41567-025-03094-2}
}

@article{kulce2021alloptical,
  title={All-optical synthesis of an arbitrary linear transformation using diffractive surfaces},
  author={Kulce, Onur and Mengu, Deniz and Rivenson, Yair and Ozcan, Aydogan},
  journal={Light: Science \& Applications},
  volume={10},
  number={1},
  pages={196},
  year={2021},
  publisher={Nature Publishing Group},
  doi={10.1038/s41377-021-00623-5}
}

@article{asgari2024theory,
  title={Theory and applications of photonic time crystals: a tutorial},
  author={Asgari, Mohammad M. and Garg, Puneet and Wang, Xuchen and Mirmoosa, Mohammad S. and Rockstuhl, Carsten and Asadchy, Viktar},
  journal={Advances in Optics and Photonics},
  volume={16},
  number={4},
  pages={958--1063},
  year={2024},
  publisher={Optica Publishing Group},
  doi={10.1364/AOP.525163}
}

@article{ciabattoni2025observation,
  title={Observation of broadband super-absorption of electromagnetic waves through space-time symmetry breaking},
  author={Ciabattoni, Matteo and Hayran, Zeki and Monticone, Francesco},
  journal={Science Advances},
  volume={11},
  number={3},
  pages={eads7407},
  year={2025},
  publisher={American Association for the Advancement of Science},
  doi={10.1126/sciadv.ads7407}
}

@article{koufidis2025chirality,
  title={Chirality-driven all-optical image differentiation},
  author={Koufidis, Stefanos Fr. and Hayran, Zeki and Monticone, Francesco and Pendry, John B. and McCall, Martin W.},
  journal={Nanophotonics},
  volume={14},
  number={27},
  pages={5449--5464},
  year={2025},
  publisher={De Gruyter},
  doi={10.1515/nanoph-2025-0479}
}

@article{hayran2026space,
  title={Space-time refraction of space-time wave packets},
  author={Hayran, Zeki and Pendry, John B.},
  journal={Advanced Photonics},
  volume={8},
  number={6},
  pages={066001},
  year={2026},
  publisher={SPIE and CLP},
  doi={10.1117/1.AP.8.6.066001}
}

@article{harwood2026programmable,
  title={Programmable Synthetic Motion at a Time-Varying Interface},
  author={Harwood, A. C. and Cielecki, D. and Raziman, T. V. and Maier, S. A. and Vezzoli, S. and Sapienza, R.},
  journal={arXiv preprint arXiv:2606.13557},
  year={2026}
}

@article{hayran2022hbar,
  title={$\hbar\omega$ versus $\hbar k$: dispersion and energy constraints on time-varying photonic materials and time crystals},
  author={Hayran, Zeki and Khurgin, Jacob B. and Monticone, Francesco},
  journal={Optical Materials Express},
  volume={12},
  number={10},
  pages={3904--3917},
  year={2022},
  publisher={Optica Publishing Group},
  doi={10.1364/OME.471672}
}

@inproceedings{kingma2015adam,
  title={Adam: A Method for Stochastic Optimization},
  author={Kingma, Diederik P. and Ba, Jimmy},
  booktitle={International Conference on Learning Representations},
  year={2015}
}

@article{lowdin1950nonorthogonality,
  title={On the non-orthogonality problem connected with the use of atomic wave functions in the theory of molecules and crystals},
  author={L{\"o}wdin, Per-Olov},
  journal={The Journal of Chemical Physics},
  volume={18},
  number={3},
  pages={365--375},
  year={1950},
  publisher={American Institute of Physics},
  doi={10.1063/1.1747632}
}

@inproceedings{mazur2019optical,
  title={Optical arbitrary waveform generator based on time-domain multiplane light conversion},
  author={Mazur, Mikael and Fontaine, Nicolas K. and Ryf, Roland and Neilson, David T. and Chen, Haoshuo and Raybon, Greg and Adamiecki, Andrew and Corteselli, Steve and Schr{\"o}der, Jochen},
  booktitle={Optical Fiber Communication Conference (OFC) 2019},
  pages={M1B.3},
  year={2019},
  organization={Optica Publishing Group},
  doi={10.1364/OFC.2019.M1B.3}
}

@article{magnus1954exponential,
  title={On the exponential solution of differential equations for a linear operator},
  author={Magnus, Wilhelm},
  journal={Communications on Pure and Applied Mathematics},
  volume={7},
  number={4},
  pages={649--673},
  year={1954},
  publisher={Wiley},
  doi={10.1002/cpa.3160070404}
}

@article{piggott2015inverse,
  title={Inverse design and demonstration of a compact and broadband on-chip wavelength demultiplexer},
  author={Piggott, Alexander Y. and Lu, Jesse and Lagoudakis, Konstantinos G. and Petykiewicz, Jan and Babinec, Thomas M. and Vu{\v{c}}kovi{\'c}, Jelena},
  journal={Nature Photonics},
  volume={9},
  number={6},
  pages={374--377},
  year={2015},
  publisher={Nature Publishing Group},
  doi={10.1038/nphoton.2015.69}
}

@article{spall1992multivariate,
  title={Multivariate stochastic approximation using a simultaneous perturbation gradient approximation},
  author={Spall, James C.},
  journal={IEEE Transactions on Automatic Control},
  volume={37},
  number={3},
  pages={332--341},
  year={1992},
  publisher={IEEE},
  doi={10.1109/9.119632}
}

\clearpage

\setcounter{section}{0}
\setcounter{subsection}{0}
\setcounter{subsubsection}{0}
\setcounter{equation}{0}
\setcounter{figure}{0}
\setcounter{table}{0}
\setcounter{footnote}{0}

\renewcommand{\thesection}{S\arabic{section}}
\renewcommand{\theequation}{S\arabic{equation}}
\renewcommand{\thefigure}{S\arabic{figure}}
\renewcommand{\thetable}{S\arabic{table}}
\renewcommand{\thefootnote}{S\arabic{footnote}}

\renewcommand{\theHsection}{supp.section.\arabic{section}}
\renewcommand{\theHequation}{supp.equation.\arabic{equation}}
\renewcommand{\theHfigure}{supp.figure.\arabic{figure}}
\renewcommand{\theHtable}{supp.table.\arabic{table}}
\renewcommand{\theHfootnote}{supp.footnote.\arabic{footnote}}

\begin{center}
{\Large\bfseries Supplementary Information\par}
\vspace{0.8em}
{\large\bfseries Reaching the optical propagation limit\\
in temporal analog computing\par}
\vspace{0.8em}
Zeki Hayran\\
The Blackett Laboratory, Department of Physics, Imperial College London, London SW7 2AZ, United Kingdom
\end{center}
\vspace{1em}

\section{Exact two-time Maxwell formulation}
\label{sec:supp_two_time_maxwell}

The two-time representation is an invertible transformation of the ordinary propagation coordinate \(z\) and laboratory time \(t\). Introducing a reference velocity \(v_r\), with \(\beta_r=v_r^{-1}\), we define
\begin{equation}
T_1=\beta_r(z-z_0),
\qquad
T_2=t-\beta_r(z-z_0),
\label{eq:supp_coordinates}
\end{equation}
where \(z_0\) denotes the input plane. The inverse transformation is
\begin{equation}
z=z_0+v_rT_1,
\qquad
t=T_1+T_2.
\label{eq:supp_inverse_coordinates}
\end{equation}
Thus \(T_1\) measures propagation distance in temporal units, while \(T_2\) is retarded physical time relative to a trajectory moving at \(v_r\). The transformation introduces no additional physical time.

The corresponding derivatives follow directly from the chain rule,
\begin{equation}
\partial_z
=
\beta_r
\left(
\partial_{T_1}-\partial_{T_2}
\right),
\qquad
\partial_t
=
\partial_{T_2}.
\label{eq:supp_derivative_transform}
\end{equation}
These relations are exact and do not require an envelope or forward-propagation approximation.

For a general electromagnetic field, define
\begin{equation}
\nabla_{2t}
=
\hat{\mathbf{x}}\partial_x
+
\hat{\mathbf{y}}\partial_y
+
\hat{\mathbf{z}}\beta_r
\left(
\partial_{T_1}-\partial_{T_2}
\right).
\label{eq:supp_two_time_gradient}
\end{equation}
Maxwell equations become
\begin{equation}
\nabla_{2t}\times\mathbf{E}
=
-\partial_{T_2}\mathbf{B},
\label{eq:supp_maxwell_faraday}
\end{equation}
\begin{equation}
\nabla_{2t}\times\mathbf{H}
=
\mathbf{J}_{\mathrm f}
+
\partial_{T_2}\mathbf{D},
\label{eq:supp_maxwell_ampere}
\end{equation}
together with
\begin{equation}
\nabla_{2t}\cdot\mathbf{B}=0,
\qquad
\nabla_{2t}\cdot\mathbf{D}=\rho_{\mathrm f}.
\label{eq:supp_maxwell_divergence}
\end{equation}
The full electromagnetic dynamics are therefore retained on the \((T_1,T_2)\) plane.

For the one-dimensional geometry used for the full-wave calculations,
\begin{equation}
\mathbf{E}
=
\hat{\mathbf{x}}E(z,t),
\qquad
\mathbf{H}
=
\hat{\mathbf{y}}H(z,t),
\end{equation}
with no free sources. The laboratory-frame Maxwell equations are
\begin{equation}
\partial_zE=-\partial_tB,
\qquad
\partial_zH=-\partial_tD.
\label{eq:supp_1d_lab_maxwell}
\end{equation}
Using Eq.~\ref{eq:supp_derivative_transform} gives
\begin{equation}
\partial_{T_1}E
=
\partial_{T_2}E
-
v_r\partial_{T_2}B,
\label{eq:supp_1d_E}
\end{equation}
\begin{equation}
\partial_{T_1}H
=
\partial_{T_2}H
-
v_r\partial_{T_2}D.
\label{eq:supp_1d_H}
\end{equation}
For a nonmagnetic medium, \(B=\mu_0H\), and elimination of \(H\) yields
\begin{equation}
\beta_r^2
\left(
\partial_{T_1}-\partial_{T_2}
\right)^2E
-
\mu_0\partial_{T_2}^2D
=
0.
\label{eq:supp_exact_wave}
\end{equation}
Equation~\ref{eq:supp_exact_wave} contains both a second derivative along \(T_1\) and a mixed \(T_1,T_2\) derivative. The first-order propagation equation used in the main text therefore emerges only after selecting a forward propagation branch.

To identify this branch, consider a monochromatic field
\begin{equation}
E(z,t)
\propto
\exp
\left[
ik(\omega)(z-z_0)-i\omega t
\right].
\end{equation}
In the two-time coordinates,
\begin{equation}
E(T_1,T_2)
\propto
\exp
\left[
-i\Omega_1(\omega)T_1-i\omega T_2
\right],
\end{equation}
where
\begin{equation}
\Omega_1(\omega)
=
\omega-v_rk(\omega).
\label{eq:supp_two_time_dispersion}
\end{equation}
For a reciprocal backward-propagating branch, \(k\rightarrow-k\), giving
\begin{equation}
\Omega_1^{(-)}(\omega)
=
\omega+v_rk(\omega).
\end{equation}

A pulse centred at frequency \(\omega\), launched at \(T_2=T_{2,0}\), follows
\begin{equation}
T_2(T_1)
=
T_{2,0}
+
\left[
v_r\beta_1(\omega)-1
\right]T_1,
\label{eq:supp_pulse_trajectory}
\end{equation}
where
\begin{equation}
\beta_1(\omega)
=
\frac{dk}{d\omega}
=
\frac{1}{v_g(\omega)}.
\end{equation}
Choosing
\begin{equation}
v_r=v_g(\omega_0)
\label{eq:supp_reference_velocity}
\end{equation}
makes the centre of the reference-frequency pulse stationary along \(T_2\) to first order. The total advance along \(T_1\) through a medium of length \(L\) is then
\begin{equation}
\Delta T_1
=
\beta_rL
=
\frac{L}{v_g(\omega_0)}
\equiv
\tau_{\mathrm{prop}}.
\label{eq:supp_T1_equals_transit}
\end{equation}
The evolution coordinate of the two-time description therefore spans the optical transit time through the computing medium.

\section{Dispersive and time-varying constitutive response}
\label{sec:supp_constitutive}

Material dispersion introduces memory along physical time and should be separated from the two coordinates used to describe propagation. For a spatially local, linear and causal dielectric, the constitutive response can be written as \cite{solis2021functional,hayran2022hbar,koutserimpas2024time}
\begin{equation}
D(z,t)
=
\epsilon_0\epsilon_\infty E(z,t)
+
\epsilon_0
\int_0^\infty
\chi(z,t;\tau_m)
E(z,t-\tau_m)
\,d\tau_m,
\label{eq:supp_causal_constitutive_lab}
\end{equation}
where \(\tau_m\geq0\) is the material-memory delay. Transforming to the coordinates of Eq.~\ref{eq:supp_coordinates} gives
\begin{equation}
D(T_1,T_2)
=
\epsilon_0\epsilon_\infty E(T_1,T_2)
+
\epsilon_0
\int_0^\infty
\chi_{2t}(T_1,T_2;\tau_m)
E(T_1,T_2-\tau_m)
\,d\tau_m.
\label{eq:supp_causal_constitutive_2t}
\end{equation}
At fixed material position, \(T_1\) is fixed. Material memory therefore connects earlier values of \(T_2\) at the same \(T_1\).

Equivalently, a general nonstationary susceptibility may be written as
\begin{equation}
D(z,t)
=
\epsilon_0\epsilon_\infty E(z,t)
+
\epsilon_0
\int_{-\infty}^{t}
\chi(z;t,t^{\prime})
E(z,t^{\prime})
\,dt^{\prime}.
\label{eq:supp_t_tprime}
\end{equation}
At fixed \(z\),
\begin{equation}
t=T_1+T_2,
\qquad
t^{\prime}=T_1+T_2^{\prime},
\end{equation}
so the corresponding two-time constitutive relation is
\begin{equation}
D(T_1,T_2)
=
\epsilon_0\epsilon_\infty E(T_1,T_2)
+
\epsilon_0
\int_{-\infty}^{T_2}
\chi_{2t}(T_1;T_2,T_2^{\prime})
E(T_1,T_2^{\prime})
\,dT_2^{\prime}.
\label{eq:supp_memory_kernel_two_time}
\end{equation}
The pair \((T_1,T_2)\) specifies the electromagnetic event, while \(T_2^{\prime}\) or \(\tau_m\) describes material memory.

For the full-wave implementation considered here, dispersion is represented by a Drude auxiliary current,
\begin{equation}
D
=
\epsilon_0\epsilon_\infty E+P,
\qquad
J=\partial_tP,
\end{equation}
which obeys
\begin{equation}
\partial_tJ
+
\gamma J
=
\epsilon_0\omega_p^2(z,t)E.
\label{eq:supp_drude_lab}
\end{equation}
Because \(\partial_t=\partial_{T_2}\), its two-time form is
\begin{equation}
\partial_{T_2}J
+
\gamma J
=
\epsilon_0
\omega_p^2(T_1,T_2)
E.
\label{eq:supp_drude_2t}
\end{equation}
We program the plasma frequency according to
\begin{equation}
\omega_p(T_1,T_2)
=
\omega_{p0}
\left[
1+\delta_p(T_1,T_2)
\right],
\label{eq:supp_wp_fraction}
\end{equation}
while keeping \(\gamma\) fixed.

For the stationary background medium, Eq.~\ref{eq:supp_drude_lab} gives
\begin{equation}
\epsilon(\omega)
=
\epsilon_\infty
-
\frac{\omega_p^2}
{\omega^2+i\gamma\omega}.
\label{eq:supp_drude_permittivity}
\end{equation}
The full-wave simulations evolve Eq.~\ref{eq:supp_drude_lab} directly rather than replacing the time-varying dispersive response by an instantaneous frequency-dependent permittivity.

\section{Forward-wave and envelope limits}
\label{sec:supp_forward_envelope}

The scalar equation used for the conceptual waveform-control demonstration follows from the exact forward branch of Eq.~\ref{eq:supp_two_time_dispersion}. Choose a carrier frequency \(\omega_0\), with
\begin{equation}
k_0=k(\omega_0),
\qquad
\Omega_{1,0}
=
\omega_0-v_rk_0,
\end{equation}
and write the analytic field as
\begin{equation}
E^{(+)}(T_1,T_2)
=
A(T_1,T_2)
\exp
\left[
-i\Omega_{1,0}T_1-i\omega_0T_2
\right].
\label{eq:supp_envelope_definition}
\end{equation}
For a frequency offset
\begin{equation}
\Omega=\omega-\omega_0,
\end{equation}
the exact unmodulated forward-envelope generator is
\begin{equation}
H_0(\Omega)
=
\Omega
-
v_r
\left[
k(\omega_0+\Omega)-k_0
\right].
\label{eq:supp_exact_envelope_generator}
\end{equation}
The envelope therefore obeys
\begin{equation}
i\partial_{T_1}A
=
H_0(\hat{\Omega})A,
\qquad
\hat{\Omega}=i\partial_{T_2}.
\label{eq:supp_exact_envelope_operator}
\end{equation}

Expanding the propagation constant around the carrier,
\begin{equation}
k(\omega_0+\Omega)
=
k_0
+
\beta_1\Omega
+
\frac{\beta_2}{2}\Omega^2
+
\frac{\beta_3}{6}\Omega^3
+\cdots,
\label{eq:supp_beta_expansion}
\end{equation}
gives
\begin{equation}
H_0(\Omega)
=
\left(
1-v_r\beta_1
\right)\Omega
-
\frac{v_r\beta_2}{2}\Omega^2
-
\frac{v_r\beta_3}{6}\Omega^3
-\cdots.
\label{eq:supp_envelope_expansion}
\end{equation}
With
\begin{equation}
v_r=\beta_1^{-1}(\omega_0),
\end{equation}
the first-order drift vanishes. Retaining quadratic dispersion gives
\begin{equation}
i\partial_{T_1}A
=
\frac{v_r\beta_2}{2}
\partial_{T_2}^2A.
\label{eq:supp_quadratic_dispersion}
\end{equation}
Defining
\begin{equation}
B_0=-v_r\beta_2
\label{eq:supp_B0}
\end{equation}
gives
\begin{equation}
i\partial_{T_1}A
=
-\frac{B_0}{2}
\partial_{T_2}^2A.
\end{equation}

A weak modulation changes the local forward propagation constant. Expanding this perturbation as
\begin{equation}
\delta k(\omega;T_1,T_2)
=
\delta k_0(T_1,T_2)
+
\delta\beta_1(T_1,T_2)\Omega
+
\frac{\delta\beta_2(T_1,T_2)}{2}\Omega^2
+\cdots,
\label{eq:supp_modulated_k}
\end{equation}
shows that its leading contribution to the propagation generator is
\begin{equation}
V(T_1,T_2)
=
-v_r\delta k_0(T_1,T_2).
\label{eq:supp_U_from_delta_k}
\end{equation}
When modulation-induced changes of group delay and higher-order dispersion are negligible over the occupied bandwidth, the envelope equation becomes
\begin{equation}
i\partial_{T_1}A
=
-\frac{B_0}{2}
\partial_{T_2}^2A
+
V(T_1,T_2)A.
\label{eq:supp_scalar_two_time}
\end{equation}

The dispersion and temporal modulation appearing in Eq.~\ref{eq:supp_scalar_two_time} are familiar ingredients of temporal waveguiding, spatiotemporal pulse coupling and temporal phase--dispersion transformations \cite{plansinis2016temporal,dong2023spatiotemporal,mazur2019optical,ashby2020temporal}. Here \(T_2\) carries the temporal function on which the operation acts, while \(T_1\) orders how the material response acting on that function evolves through the medium.

Defining
\begin{equation}
\hat{D}
=
-\frac{B_0}{2}\partial_{T_2}^2,
\qquad
\hat{V}
=
V(T_1,T_2),
\end{equation}
a short propagation interval can be factorized as
\begin{equation}
\exp
\left[
-i(\hat{D}+\hat{V})\Delta T_1
\right]
=
\exp
\left[
-\frac{i}{2}\hat{V}\Delta T_1
\right]
\exp
\left[
-i\hat{D}\Delta T_1
\right]
\exp
\left[
-\frac{i}{2}\hat{V}\Delta T_1
\right]
+
O(\Delta T_1^3).
\label{eq:supp_strang}
\end{equation}
The outer factors are temporal phase operations and the middle factor is dispersive propagation. Cascaded phase--dispersion processing is therefore recovered as an operator-split realization of this continuous-propagation limit.

The Fredholm solver includes the modulation-induced change of the local dispersive propagation law through the propagation model described in Supplementary Section~S8.

\section{Ordered temporal operator synthesis}
\label{sec:supp_operator_synthesis}

Equation~\ref{eq:supp_scalar_two_time} can be written as
\begin{equation}
i\partial_{T_1}
\left|
A(T_1)
\right\rangle
=
\hat{H}(T_1)
\left|
A(T_1)
\right\rangle,
\label{eq:supp_operator_evolution}
\end{equation}
where
\begin{equation}
\hat{H}(T_1)
=
-\frac{B_0}{2}\partial_{T_2}^2
+
V(T_1,T_2).
\label{eq:supp_H}
\end{equation}
The corresponding input--output evolution is
\begin{equation}
\hat{\mathcal{U}}
(T_{1,f},T_{1,i})
=
\mathcal{T}_1
\exp
\left[
-i
\int_{T_{1,i}}^{T_{1,f}}
\hat{H}(T_1)
\,dT_1
\right].
\label{eq:supp_ordered_U}
\end{equation}

For a material response that varies along \(T_1\), the propagation generators generally satisfy
\begin{equation}
\left[
\hat{H}(T_1),
\hat{H}(T_1^{\prime})
\right]
\neq0.
\label{eq:supp_noncommuting}
\end{equation}
The resulting transformation therefore depends on the ordering of the temporal operations during propagation.

This dependence is explicit in the Magnus representation \cite{magnus1954exponential},
\begin{equation}
\hat{\mathcal{U}}
=
\exp
\left[
\hat{\mathcal{M}}_1
+
\hat{\mathcal{M}}_2
+\cdots
\right],
\label{eq:supp_magnus}
\end{equation}
where
\begin{equation}
\hat{\mathcal{M}}_1
=
-i
\int_{T_{1,i}}^{T_{1,f}}
\hat{H}(\tau)
\,d\tau,
\label{eq:supp_magnus1}
\end{equation}
and
\begin{equation}
\hat{\mathcal{M}}_2
=
-\frac{1}{2}
\int_{T_{1,i}}^{T_{1,f}}
d\tau_1
\int_{T_{1,i}}^{\tau_1}
d\tau_2
\left[
\hat{H}(\tau_1),
\hat{H}(\tau_2)
\right].
\label{eq:supp_magnus2}
\end{equation}
The first term contains the accumulated generator, while higher-order terms contain the effects of ordered noncommuting transformations.

The complete propagation operator defines a temporal transfer kernel
\begin{equation}
S(T_2,T_2^{\prime})
=
\left\langle
T_2
\middle|
\hat{\mathcal{U}}
\middle|
T_2^{\prime}
\right\rangle,
\label{eq:supp_transfer_kernel}
\end{equation}
such that
\begin{equation}
A_{\mathrm{out}}(T_2)
=
\int_{\mathcal{T}}
S(T_2,T_2^{\prime})
A_{\mathrm{in}}(T_2^{\prime})
\,dT_2^{\prime}.
\label{eq:supp_transfer_integral}
\end{equation}
The dispersive term mixes different values of \(T_2\), while the programmed material response controls their subsequent amplitudes and phases during propagation.

The role of dispersion becomes clear in the limit \(B_0=0\). The evolution is then local in waveform time and
\begin{equation}
S(T_2,T_2^{\prime})
=
\exp
\left[
-i
\int
V(T_1,T_2)
\,dT_1
\right]
\delta(T_2-T_2^{\prime}).
\label{eq:supp_local_limit}
\end{equation}
A general nonlocal temporal operator therefore requires temporal mixing in addition to local modulation.

For a finite computational basis \(\{\phi_n\}_{n=1}^{N}\), with
\begin{equation}
\int
\phi_m^*(T_2)
\phi_n(T_2)
\,dT_2
=
\delta_{mn},
\end{equation}
the implemented matrix is
\begin{equation}
S_{mn}
=
\left\langle
\phi_m
\middle|
\hat{\mathcal{U}}
\middle|
\phi_n
\right\rangle.
\label{eq:supp_Smodal}
\end{equation}
Even when the complete electromagnetic evolution is lossless, the operator restricted to a finite computational subspace can be nonunitary if part of the field couples outside that space. Material absorption provides an additional source of nonunitarity in the Maxwell--Drude realization. These mechanisms allow the finite-dimensional propagation operator to represent nonunitary Fredholm inverses as well as unitary transformations.

\section{Synthetic-motion realization}
\label{sec:supp_synthetic_motion}

A two-time material programme corresponds to an ordinary space-time modulation when expressed in laboratory coordinates. Let \(m\) denote a local material parameter. A general two-time response can be written as
\begin{equation}
m(z,t)
=
m_{2t}
\left[
\beta_r(z-z_0),
t-\beta_r(z-z_0)
\right].
\label{eq:supp_lab_mapping}
\end{equation}

Consider first a modulation travelling at velocity \(v_m\),
\begin{equation}
m(z,t)
=
m_0
\left[
t-\beta_m(z-z_0)
\right],
\qquad
\beta_m=v_m^{-1}.
\label{eq:supp_travelling_mod}
\end{equation}
In the two-time coordinates this becomes
\begin{equation}
m(T_1,T_2)
=
m_0
\left[
T_2+
\left(
1-\frac{v_r}{v_m}
\right)T_1
\right].
\label{eq:supp_moving_transform}
\end{equation}
When
\begin{equation}
v_m=v_r,
\end{equation}
the modulation becomes stationary with respect to the waveform coordinate,
\begin{equation}
m(T_1,T_2)=m_0(T_2).
\label{eq:supp_velocity_matching}
\end{equation}
This provides the co-moving temporal structure used in the synthetic-motion interpretation of Fig.~2d \cite{huidobro2019fresnel,galiffi2022photonics,harwood2025space}.

More generally, the shape or amplitude of this co-moving modulation can change as propagation proceeds,
\begin{equation}
m=m(T_1,T_2),
\end{equation}
providing independent control across the waveform and along its propagation.

For a modulation travelling at \(v_m\), a surface of constant modulation phase obeys
\begin{equation}
\frac{dT_2}{dT_1}
=
\frac{v_r}{v_m}-1.
\label{eq:supp_modulation_slope}
\end{equation}
Similarly, an optical pulse centred at frequency \(\omega\) follows
\begin{equation}
\frac{dT_2}{dT_1}
=
\frac{v_r}{v_g(\omega)}-1.
\label{eq:supp_pulse_slope}
\end{equation}
Arrival time therefore determines the intercept of the trajectory through the two-time modulation, while carrier frequency changes its slope through material dispersion. These relations also determine the finite timing and frequency acceptance of a given programmed region.

\section{Inverse design of time-selective waveform transformations}
\label{sec:supp_waveform_design}

The two input pulses, governing propagation parameters and fidelity measure used for Fig.~3 are defined in Methods. The target output intensities are one-dimensional temporal profiles derived from silhouettes of the Palace of Westminster and Tower Bridge. Each profile is mapped onto a \(2.16~\mathrm{ps}\) temporal interval and smoothed with a \(5~\mathrm{fs}\) RMS Gaussian to impose finite temporal bandwidth. Only the output intensity is specified. The complex output phase remains unconstrained, allowing the two initially orthogonal input states to remain orthogonal even when their output intensity profiles overlap.

The modulation is represented by a smooth bounded expansion,
\begin{equation}
V(T_1,T_2)
=
V_{\max}
\tanh
\left[
E(T_1,T_2)
\sum_{\mu,\nu}
C_{\mu\nu}
B_\mu^{(1)}(T_1)
B_\nu^{(2)}(T_2)
\right],
\label{eq:supp_skyline_potential}
\end{equation}
where \(B_\mu^{(1)}\) and \(B_\nu^{(2)}\) are Gaussian radial basis functions and \(E(T_1,T_2)\) smoothly confines the modulation to the interaction region. The control surface contains \(54\times108\) real coefficients.

Propagation is evaluated using a symmetric split-step method,
\begin{equation}
A(T_1+\Delta T_1)
\simeq
e^{-i\hat{V}\Delta T_1/2}
e^{-i\hat{D}\Delta T_1}
e^{-i\hat{V}\Delta T_1/2}
A(T_1),
\label{eq:supp_skyline_split}
\end{equation}
where
\begin{equation}
\hat{D}
=
-\frac{B_0}{2}\partial_{T_2}^2.
\end{equation}
The propagation interval is \(2.1~\mathrm{ps}\), sampled with 4409 steps along \(T_1\). The \(T_2\) calculation uses 6144 samples over an \(8.4~\mathrm{ps}\) numerical window. A weak absorber outside the useful waveform region suppresses FFT wrap-around.

Using the fidelities \(S_1\) and \(S_2\) defined in Methods, the optimization objective combines their arithmetic and harmonic means,
\begin{equation}
S_{\mathrm{mean}}
=
\frac{S_1+S_2}{2},
\end{equation}
\begin{equation}
S_{\mathrm{harm}}
=
\frac{2}
{(S_1+10^{-5})^{-1}+(S_2+10^{-5})^{-1}},
\end{equation}
according to
\begin{equation}
J_{\mathrm{shape}}
=
0.45S_{\mathrm{mean}}
+
0.55S_{\mathrm{harm}}.
\label{eq:supp_skyline_balanced}
\end{equation}
Weak penalties suppress excessive control variation and spectral content outside the intended envelope band. The gradient is calculated from the adjoint of the discrete propagation operator, following standard photonic inverse-design methods \cite{piggott2015inverse}.

The final design gives
\begin{equation}
S_{\mathrm{Westminster}}
=
0.9818,
\qquad
S_{\mathrm{Tower~Bridge}}
=
0.9800,
\label{eq:supp_skyline_final_scores}
\end{equation}
with worst-case fidelity
\begin{equation}
S_{\mathrm{worst}}
=
0.9800.
\end{equation}
The retained output powers are \(0.9832\) and \(0.9806\), respectively. The magnitude of the complex overlap between the two output states is
\begin{equation}
\left|
\left\langle
A_1^{\mathrm{out}}
\middle|
A_2^{\mathrm{out}}
\right\rangle
\right|
=
1.6\times10^{-6}.
\label{eq:supp_skyline_overlap}
\end{equation}

\section{Fredholm operator in a finite temporal basis}
\label{sec:supp_fredholm}

The analytical Fredholm kernel, temporal scale, integration domain and nominal frequencies of the five computational modes are defined in Methods. The corresponding dimensional kernel is
\begin{equation}
K(T_2,T_2^{\prime})
=
\frac{1}{T_s}
\kappa
\left(
\frac{T_2}{T_s},
\frac{T_2^{\prime}}{T_s}
\right).
\label{eq:supp_dimensional_kernel}
\end{equation}

To construct the finite temporal basis, we begin from the windowed Fourier functions
\begin{equation}
\phi_n^{\mathrm{raw}}(T_2)
=
w(T_2)
\exp
\left(
-i2\pi f_nT_2
\right),
\qquad
n=-2,\ldots,2,
\label{eq:supp_raw_basis}
\end{equation}
where
\begin{equation}
w(T_2)
=
\exp
\left[
-\frac{1}{2}
\left(
\frac{|T_2|}{500~\mathrm{fs}}
\right)^8
\right].
\label{eq:supp_basis_window}
\end{equation}
The common temporal window makes the raw modes slightly nonorthogonal. We therefore apply the symmetric L\"owdin orthonormalization stated in Methods. If \(\Phi_{\mathrm{raw}}\) contains the normalized raw basis functions and
\begin{equation}
G
=
\Delta T_2
\Phi_{\mathrm{raw}}^\dagger
\Phi_{\mathrm{raw}},
\label{eq:supp_gram_matrix}
\end{equation}
the orthonormal basis is
\begin{equation}
\Phi
=
\Phi_{\mathrm{raw}}
G^{-1/2},
\label{eq:supp_lowdin}
\end{equation}
which satisfies
\begin{equation}
\Delta T_2
\Phi^\dagger\Phi
=
I_5.
\label{eq:supp_discrete_orthonormality}
\end{equation}

The modal representation of the continuous Fredholm kernel is
\begin{equation}
K_{mn}^{(5)}
=
\int_{\mathcal{T}}dT_2
\int_{\mathcal{T}}dT_2^{\prime}
\,
\phi_m^*(T_2)
K(T_2,T_2^{\prime})
\phi_n(T_2^{\prime}).
\label{eq:supp_Kmodal_continuous}
\end{equation}
The projected matrix \(K^{(5)}\) defines the target solution operator given in Methods. For the selected Fredholm problem,
\begin{equation}
\mathrm{cond}
\left(
I_5-K^{(5)}
\right)
=
1.146,
\label{eq:supp_condition}
\end{equation}
so the inversion is well conditioned within the selected temporal computational space.

Conversely, a physical propagation matrix \(S_{\mathrm{physical}}^{(5)}\) defines an implemented Fredholm kernel through
\begin{equation}
K_{\mathrm{physical}}^{(5)}
=
I_5-
\left(
S_{\mathrm{physical}}^{(5)}
\right)^{-1}.
\label{eq:supp_Kphysical}
\end{equation}
This is the implemented matrix compared with the target in Fig.~4b.

The continuous analytical kernel therefore defines the target equation through its projection onto the five-mode temporal space. The reported kernel, propagation-operator and solution errors refer to this computational subspace.

\section{Physics-constrained inverse design in a Maxwell--Drude medium}
\label{sec:supp_inverse_design}

The physical parameters of the Maxwell--Drude system and the allowed plasma-frequency modulation range are given in Methods. These parameters are related to the numerically equivalent inverse-design problem by the similarity transformation
\begin{equation}
z\rightarrow sz,
\qquad
t\rightarrow st,
\qquad
\omega\rightarrow\frac{\omega}{s},
\label{eq:supp_similarity}
\end{equation}
with
\begin{equation}
s=5.
\end{equation}
The Drude parameters transform as
\begin{equation}
\omega_p\rightarrow\frac{\omega_p}{s},
\qquad
\gamma\rightarrow\frac{\gamma}{s},
\end{equation}
while \(\epsilon_\infty\) and the fractional plasma-frequency modulation remain unchanged. For the Drude response defined in Supplementary Section~S2,
\begin{equation}
\epsilon_{\mathrm{scaled}}
\left(
\frac{\omega}{s}
\right)
=
\epsilon(\omega),
\label{eq:supp_similarity_permittivity}
\end{equation}
so the dimensionless Maxwell--Drude propagation problem and its modal operator are unchanged. Applying this transformation gives the physical parameters reported in Methods. The corresponding carrier and background plasma frequencies are approximately
\begin{equation}
\frac{\omega_0}{2\pi}=38.68~\mathrm{THz},
\qquad
\frac{\omega_{p0}}{2\pi}=61.89~\mathrm{THz}.
\label{eq:supp_physical_frequencies}
\end{equation}

The material modulation is written as
\begin{equation}
\omega_p(T_1,T_2)
=
\omega_{p0}
\left[
1+\delta_p(T_1,T_2)
\right],
\label{eq:supp_wp_program}
\end{equation}
with \(\delta_p(T_1,T_2)\) constrained to the modulation range stated in Methods.

All matrices in this section refer to the five-mode projection defined in Supplementary Section~S7. The inverse design is constrained directly by the physical propagation dynamics and the accessible material response. Each candidate plasma-frequency profile is evaluated through the dispersive Drude propagation model, so the modulation-induced change of local material dispersion enters the optimization together with the desired modal transformation. Leakage outside the computational space, departure from the initial physical profile and excessive modulation are incorporated into the objective.

The optimization is initialized from a smooth modulation profile with a maximum plasma-frequency reduction of \(4.5\%\). The optimized correction is represented by 35 smooth real control parameters distributed over propagation and waveform time. A bounded logistic mapping enforces the physical modulation constraint throughout the optimization.

Furthermore, we retain the change of local material dispersion produced by \(\delta_p(T_1,T_2)\). Let
\begin{equation}
k_{\mathrm{bg}}(\Omega)
=
k(\omega_0+\Omega;\omega_{p0})
\label{eq:supp_kbg}
\end{equation}
denote the background propagation constant, and define
\begin{equation}
D_{\mathrm{bg}}(\Omega)
=
k_{\mathrm{bg}}(\Omega)
-
k_0
-
\beta_1\Omega.
\label{eq:supp_Dbg}
\end{equation}
For a local fractional plasma-frequency shift \(\delta_p\),
\begin{equation}
\Delta k(\Omega;\delta_p)
=
k
\left[
\omega_0+\Omega;
\omega_{p0}(1+\delta_p)
\right]
-
k_{\mathrm{bg}}(\Omega).
\label{eq:supp_delta_k_local}
\end{equation}
The dependence of this local dispersive response on \(\delta_p\) and \(\Omega\) is represented in separable form as
\begin{equation}
\Delta k(\Omega;\delta_p)
\simeq
\sum_r
a_r(\delta_p)b_r(\Omega),
\label{eq:supp_local_svd}
\end{equation}
obtained by singular-value decomposition over the accessible modulation and spectral ranges. Terms are retained until the relative truncation error falls below \(10^{-6}\), with at most five components.

When the plasma-frequency modulation varies across waveform time, the local dispersive operator at each \(T_1\) is represented symmetrically as
\begin{equation}
\hat{D}_{\mathrm{loc}}(T_1)
=
\frac{1}{2}
\sum_r
\left[
a_r\!\left(\delta_p(T_1,T_2)\right)b_r(\hat{\Omega})
+
b_r(\hat{\Omega})a_r\!\left(\delta_p(T_1,T_2)\right)
\right].
\label{eq:supp_symmetric_dispersion}
\end{equation}
The complete forward-envelope evolution used in the inverse design can therefore be written as
\begin{equation}
i\partial_z A(z,T_2)
=
-
\left[
\hat{D}_{\mathrm{bg}}
+
\hat{D}_{\mathrm{loc}}(z)
\right]
A(z,T_2),
\label{eq:supp_fredholm_forward_model}
\end{equation}
where \(\hat{D}_{\mathrm{bg}}=D_{\mathrm{bg}}(\hat{\Omega})\), and the dependence on \(z\) is equivalently parameterized by \(T_1=\beta_r(z-z_0)\). Numerically, Eq.~\ref{eq:supp_fredholm_forward_model} is propagated using symmetric splitting between the background and local dispersive operators. The forward propagation model therefore retains both the background dispersion and its local modification by the programmed material response.

The target is approached through continuation from the initial physical operator. Let \(K_{\mathrm{seed}}\) denote the Fredholm kernel reconstructed from the initial physical modulation profile. We then define
\begin{equation}
K_\alpha
=
(1-\alpha)K_{\mathrm{seed}}
+
\alpha K_{\mathrm{target}},
\qquad
\alpha\in
\{
0.25,0.50,0.75,1
\},
\label{eq:supp_continuation}
\end{equation}
with the corresponding solution operator
\begin{equation}
S_\alpha
=
\left(
I-K_\alpha
\right)^{-1}.
\label{eq:supp_continuation_solution}
\end{equation}

For a calculated propagation matrix \(S\), the equation residual is defined as
\begin{equation}
\epsilon_{\mathrm{eq}}
=
\frac{
\|
(I-K_{\mathrm{target}})S-I
\|_F
}
{\sqrt{5}},
\label{eq:supp_equation_residual}
\end{equation}
while the propagation-operator error is
\begin{equation}
\epsilon_S
=
\frac{
\|
S-S_{\mathrm{target}}
\|_F
}
{
\|
S_{\mathrm{target}}
\|_F
}.
\label{eq:supp_operator_error}
\end{equation}
Let \(\Psi_{\mathrm{out}}\) denote the matrix whose columns are the full propagated output fields obtained by exciting each computational basis mode. Leakage outside the computational space is then quantified by
\begin{equation}
\eta_{\mathrm{leak}}
=
\frac{
\|
\Psi_{\mathrm{out}}-\Phi S
\|_F^2
}
{
\|
\Psi_{\mathrm{out}}
\|_F^2
}.
\label{eq:supp_leakage}
\end{equation}

The optimization objective is
\begin{equation}
J
=
\epsilon_{\mathrm{eq}}^2
+
0.25\eta_{\mathrm{leak}}
+
0.012\epsilon_{\mathrm{seed}}^2
+
0.0015P_{\mathrm{ctrl}}
+
0.020P_{\mathrm{peak}}^2,
\label{eq:supp_inverse_objective}
\end{equation}
where \(\epsilon_{\mathrm{seed}}\) measures the relative change from the initial modulation, \(P_{\mathrm{ctrl}}\) penalizes the magnitude of the control parameters and \(P_{\mathrm{peak}}\) penalizes plasma-frequency modulation amplitudes above the preferred range.

The gradient is estimated using simultaneous perturbation stochastic approximation \cite{spall1992multivariate}, and the controls are updated using Adam \cite{kingma2015adam}. The optimization is performed over the continuation sequence of Eq.~\ref{eq:supp_continuation}, after which the final material profile is evaluated with a higher-resolution propagation calculation.

The optimized modulation reaches
\begin{equation}
\max
\left|
\frac{\Delta\omega_p}{\omega_{p0}}
\right|
=
4.494\%.
\label{eq:supp_final_modulation}
\end{equation}
The corresponding five-mode propagation operator gives a normalized complex kernel overlap
\begin{equation}
\mathcal{O}_K
=
\frac{
\left|
\mathrm{Tr}
\left(
K_{\mathrm{target}}^\dagger K_{\mathrm{physical}}
\right)
\right|
}{
\|K_{\mathrm{target}}\|_F
\|K_{\mathrm{physical}}\|_F
}
=
0.9985.
\label{eq:supp_kernel_overlap}
\end{equation}
The relative kernel error is
\begin{equation}
\epsilon_K
=
\frac{
\|
K_{\mathrm{physical}}-K_{\mathrm{target}}
\|_F
}{
\|
K_{\mathrm{target}}
\|_F
}
=
6.05\%,
\label{eq:supp_final_kernel_error}
\end{equation}
with corresponding solution-operator and equation residuals
\begin{equation}
\epsilon_S
=
10.80\%,
\qquad
\epsilon_{\mathrm{eq}}
=
10.99\%.
\label{eq:supp_final_operator_errors}
\end{equation}
The total output power outside the five-mode computational space is
\begin{equation}
\eta_{\mathrm{leak}}
=
7.38\%.
\label{eq:supp_final_leakage}
\end{equation}
For the representative input used in Fig.~4, the projected propagation-model solution differs from the ideal Fredholm solution by
\begin{equation}
\epsilon_{\mathrm{proj}}
=
7.90\%.
\label{eq:supp_projected_solution_error}
\end{equation}

\section{Full-wave validation}
\label{sec:supp_fdtd}

The inverse-designed medium is validated independently using one-dimensional Maxwell--Drude finite-difference time-domain calculations. The fields obey
\begin{equation}
\partial_zE
=
-\mu_0\partial_tH,
\label{eq:supp_fdtd_faraday}
\end{equation}
\begin{equation}
\partial_zH
=
-\epsilon_0\epsilon_\infty\partial_tE
-
J,
\label{eq:supp_fdtd_ampere}
\end{equation}
together with
\begin{equation}
\partial_tJ
+
\gamma J
=
\epsilon_0
\omega_p^2(z,t)
E.
\label{eq:supp_fdtd_drude}
\end{equation}

The optimized material profile is stored using the normalized propagation coordinate
\begin{equation}
\xi
=
\frac{T_1}{\Delta T_1}
=
\frac{z-z_0}{L}.
\label{eq:supp_fdtd_xi}
\end{equation}
We therefore define the normalized representation of the optimized two-time modulation as
\begin{equation}
\delta_{p,\xi}(\xi,T_2)
=
\delta_p(\xi\Delta T_1,T_2),
\label{eq:supp_fdtd_normalized_modulation}
\end{equation}
where \(\delta_p(T_1,T_2)\) is the modulation defined in Supplementary Section~S8. For the group-velocity-matched reference frame, the corresponding laboratory-frame plasma frequency is
\begin{equation}
\omega_p(z,t)
=
\omega_{p0}
\left[
1+
\delta_{p,\xi}
\left(
\frac{z-z_0}{L},
t-\beta_1(z-z_0)
\right)
\right].
\label{eq:supp_fdtd_program}
\end{equation}
Here \(\beta_1=v_g^{-1}(\omega_0)=\beta_r\).

The equations are integrated on a staggered Yee grid using an auxiliary differential equation for the Drude current. The spatial resolution is set to at least 24 cells per shortest relevant wavelength, with Courant factor \(0.93\). Seventy-one field monitors sample the evolution through the computing region.

The incident waveform is synthesized from the same five temporal modes used to define the Fredholm equation and initialized on the forward electromagnetic branch. An otherwise identical unmodulated calculation provides a reference for the weak background attenuation and phase accumulated in the dispersive Drude medium.

At each monitor, the optical field is converted to a complex envelope around the carrier and expressed in the co-moving coordinate
\begin{equation}
T_2
=
t-t_{\mathrm{in}}
-
\beta_1z_{\mathrm{rel}}.
\label{eq:supp_fdtd_T2}
\end{equation}
The real and imaginary components shown in Fig.~4c are therefore the two quadratures of the complex optical envelope.

The extracted input and output envelopes are projected onto the computational basis,
\begin{equation}
\mathbf{c}_{\mathrm{in}}^{\mathrm{FDTD}}
=
\Delta T_2
\Phi^\dagger
A_{\mathrm{in}}^{\mathrm{FDTD}},
\label{eq:supp_fdtd_cin}
\end{equation}
\begin{equation}
\mathbf{c}_{\mathrm{out}}^{\mathrm{FDTD}}
=
\Delta T_2
\Phi^\dagger
A_{\mathrm{out}}^{\mathrm{FDTD}}.
\label{eq:supp_fdtd_cout}
\end{equation}
The projected full-wave output is
\begin{equation}
A_{\mathrm{out,proj}}^{\mathrm{FDTD}}
=
\Phi
\mathbf{c}_{\mathrm{out}}^{\mathrm{FDTD}}.
\label{eq:supp_fdtd_projection}
\end{equation}

For the equation-level comparison, the ideal Fredholm solution is evaluated using the modal coefficients present at the FDTD entrance,
\begin{equation}
A_{\mathrm{target}}^{\mathrm{FDTD}}
=
\Phi
S_{\mathrm{target}}
\mathbf{c}_{\mathrm{in}}^{\mathrm{FDTD}}.
\label{eq:supp_target_from_fdtd}
\end{equation}

Using the relative field error and normalized overlap defined in Methods, the projected full-wave output differs from the ideal Fredholm solution by
\begin{equation}
\epsilon_{\mathrm{FDTD,target}}
=
8.94\%,
\label{eq:supp_fdtd_target_error}
\end{equation}
with overlap
\begin{equation}
\mathcal{O}_{\mathrm{FDTD,target}}
=
0.9974.
\label{eq:supp_fdtd_target_overlap}
\end{equation}

The raw Maxwell--Drude output and the independently calculated forward propagation model differ by
\begin{equation}
\epsilon_{\mathrm{FDTD,prop}}
=
9.34\%,
\label{eq:supp_fdtd_prop_error}
\end{equation}
with overlap
\begin{equation}
\mathcal{O}_{\mathrm{FDTD,prop}}
=
0.9957.
\label{eq:supp_fdtd_prop_overlap}
\end{equation}

The component of the full-wave output outside the selected temporal subspace is quantified by the field-projection error
\begin{equation}
\epsilon_{\mathrm{space}}
=
\frac{
\|
A_{\mathrm{out}}^{\mathrm{FDTD}}
-
A_{\mathrm{out,proj}}^{\mathrm{FDTD}}
\|_2
}{
\|
A_{\mathrm{out}}^{\mathrm{FDTD}}
\|_2
}
=
25.17\%.
\label{eq:supp_fdtd_projection_error}
\end{equation}
Figure~4c therefore displays the component of the Maxwell--Drude evolution within the computational space used to define the Fredholm equation.

The operator comparison and the full-wave comparison test complementary aspects of the solver. The complete \(5\times5\) implemented operator in Fig.~4b is obtained by propagating the five basis inputs through the forward dispersive model. The Maxwell--Drude simulation independently tests the corresponding solution for a nontrivial superposition of these modes.

\end{document}